%% file: ms_08_11_11.tex
\documentclass[12pt,preprint]{aastex}
\def\deg{\hbox{$^\circ$}}
\newcommand{\etal}{{\it et al. }}
\newcommand{\asec}{$^{\prime\prime }$}
\newcommand{\amin}{$^\prime $}
\shortauthors{Erickson et al.}
\shorttitle{The Initial Mass Function of the $\rho$ Ophiuchi Cloud}

\begin{document}

\title{The Initial Mass Function and Disk Frequency of the $\rho$ Ophiuchi Cloud:
An Extinction-Limited Sample}

\author{Kristen L. Erickson \altaffilmark{1}}
\affil{Department of Physics and Astronomy, University of Missouri-St.  Louis\\
1 University Boulevard, St.  Louis, MO 63121\\ kle6x2@mail.umsl.edu}

\author{Bruce A. Wilking \altaffilmark{1}}
\affil{Department of Physics and Astronomy, University of Missouri-St.  Louis\\
1 University Boulevard, St.  Louis, MO 63121\\ bwilking@umsl.edu}

\author{Michael R. Meyer}
\affil{Institute for Astronomy Swiss Federal Institute of Technology Wolfgang-Pauli-Strasse 27, CH-8093 Zurich Switzerland \\ mmeyer@phys.ethz.ch}

\author{John G. Robinson \altaffilmark{2}}
\affil{Department of Physics and Astronomy, University of Missouri-St.  Louis\\
1 University Boulevard, St.  Louis, MO 63121\\ jrobinson@gmail.com}

\and

\author{Lauren N. Stephenson \altaffilmark{1}}
\affil{Department of Physics and Astronomy, University of Missouri-St.  Louis\\
1 University Boulevard, St.  Louis, MO 63121\\ lnafff@mail.umsl.edu}

\altaffiltext{1}{Visiting Astronomer, Kitt Peak National Observatory, 
National Optical Astronomy Observatory, which is operated by the Association of 
Universities for Research in Astronomy (AURA) under cooperative agreement with the National Science Foundation.}

\altaffiltext{2}{Visiting astronomer, Cerro Tololo Inter-American Observatory, 
National Optical Astronomy Observatory, which are operated by the Association of 
Universities for Research in Astronomy, under contract with the National Science Foundation.}

\begin{abstract}
We have completed an optical spectroscopic survey of an unbiased, extinction-limited sample of candidate 
young stars covering 1.3 deg$^2$ of the $\rho$ Ophiuchi star forming region. While infrared, X-ray, and optical 
surveys of the cloud have identified many young stellar objects (YSOs), these surveys 
are biased towards
particular stages of stellar evolution and are not optimal for studies of the disk frequency
and initial mass function.
We have obtained over 300 optical spectra to help identify 
135 association members based on the presence of H$\alpha$ in emission, lithium absorption, X-ray emission, 
a mid-infrared excess, a common proper motion, reflection nebulosity, and/or extinction considerations. 
Spectral types along with R and I band photometry were used to derive effective temperatures 
and bolometric luminosities for association members to compare with
theoretical tracks and isochrones for pre-main-sequence stars. 
An average age of 3.1 Myr is derived for this population which is intermediate between that 
of objects embedded in the cloud core of $\rho$ Ophiuchi and low mass stars in the Upper Scorpius subgroup. 
Consistent with this age we find a circumstellar disk frequency of  27\% $\pm$ 5\%.  
We also constructed an initial mass function for an extinction-limited sample of 123 YSOs
(A$_v$$\le$ 8 mag), which is consistent with the field star initial mass function for 
YSOs with masses $>$0.2 M$_{\sun}$. 
There may be a deficit of brown dwarfs but this result relies on completeness corrections 
and requires confirmation. 

\end{abstract}

\section{Introduction}

An important question in the theory of star formation is whether the initial mass 
function (IMF) of stars is universal. Variations in the IMF from region to region may hold clues to the roles of 
accretion, fragmentation, and ejection in producing the stellar mass spectrum (e.g., Bonnell \etal
2007).
One of the best places to investigate the 
IMF is in molecular clouds with active star formation since cluster membership is well-determined, 
low mass stars have had a limited amount of time to segregate,
and one can associate variations in the IMF to the physical conditions of the cloud.
Due to large columns of dust obscuring all but the brightest objects in the cloud,
IMF studies of young clusters require unbiased, extinction
limited spectroscopic surveys (e.g., Bastian \etal 2010).

The $\rho$ Ophiuchi molecular cloud complex is a well-studied, nearby 
region of active star formation (see Wilking \etal 2008 for review). Located at
130 pc from the Sun (Mamajek 2008), its proximity guarantees access to the broadest
range of luminosity and mass.  Most recently,
the Spitzer Space Telescope surveyed $\rho$ Ophiuchi in the mid- and far-infrared 
as part of the Legacy and guaranteed time programs. About 292 young stellar objects (YSOs) were identified
with infrared excesses due to circumstellar disks (Evans \etal 2009) over a field of view of 6.8 square degrees.   
X-ray studies have also been conducted of the region with ROSAT, XMM-Newton,
ASCA, and Chandra revealing more evolved YSOs with magnetic surface activity (Grosso \etal 2000; Gagn\'e \etal 
2004; Ozawa \etal 2005). 
But for studies of the IMF, one must sample all phases of pre-main-sequence (PMS)
evolution from heavily embedded YSOs in their main accretion phase, to classical T Tauri stars (CTTS), to 
weak-emission T Tauri stars (WTTS) with little or no circumstellar dust.
Optical spectroscopic 
surveys targeting CTTS and WTTS have been conducted of this region, however these studies 
have been biased toward objects with X-ray emission or YSOs with
suspected H$\alpha$ emission (Bouvier \& Appenzeller 1992; Martin \etal 1998; Wilking \etal 2005).
Wilking \etal (2005, hereafter Paper I) obtained 136 spectra from 5820\AA - 8700\AA\ 
at a resolution of 2.9 \AA\ and identified 88 cluster members in the main L~1688 cloud of the Ophiuchus complex.
The members had a median age of 2.1 Myr and included 
39 CTTS.  However, their survey had a selection bias toward YSOs with H$\alpha$ emission.  

In this paper, we present the results of a new optical spectroscopic survey which, when combined
with data from Paper I, enables us to construct 
an unbiased, extinction-limited sample of YSOs in the L~1688 cloud.
Section 2 describes this new spectroscopic survey which covered the wavelength range 
6249\AA -7657\AA\ with a resolution of 1.4 \AA\ that enabled us to resolve Li in absorption, an indicator of youth.
The analysis of the spectra to derive spectral types and exclude background giants is described briefly in Section 3.
Section 4 discusses the results of our analysis including the identification of association members, their spatial
distribution, their placement in a Hertzsprung-Russell (H-R) diagram relative to several theoretical models and their
age distribution.
The Section concludes by defining an extinction-limited sample and the resulting disk frequency and
distribution of masses in this subsample for comparisons with other star-forming regions.  Finally, Section 5 
compares low-extinction YSOs in L~1688 with those in Upper Scorpius and explores the relationship between star formation in Upper Scorpius, the surface population in L~1688, and that in the cloud core.

\section{Observations and Data Reduction}

Over 200 moderate resolution spectra were obtained for 184 stars identified through 
R and I band photometry as candidate YSOs.  Fifty-two of these stars had 
spectral classifications reported in Paper I, allowing us to refine our previous spectral 
classifications. Newly observed YSO candidates numbered 133.  
These observations are described in detail in the following sections. 

\subsection{Sample Selection}

Candidate YSOs were selected from an I vs. (R-I) color magnitude diagram.  
R and I band photometry were obtained from short (5 minute) exposure images 
obtained with the 0.6 m Curtis-Schmidt Telescope located at Cerro Tololo Inter-American 
Observatory in 1995 March (see also Wilking \etal 1997 for description). 
The CCD images covered a 67\amin\ x 68.5\amin\ area centered on RA(2000) = 
16$^h$ 27$^m$ 14.7$^s$, decl.(2000) = -24\deg 30\amin 06\asec\ 
with a scale of 2.03\asec\ pixel$^{-1}$.  The survey area is shown in Figure 1 relative to
the distribution of molecular gas.
Photometry was performed using an 8.1\asec\ diameter 
aperture optimized for the 4 \asec\ full width at half-maximum of the point-spread function 
(Howell 1989) with local sky background measured in an annulus 10 - 20 \asec\ in diameter.  
Zero-points were computed in the Kron-Cousins photometric system using standard star fields 
established by Landolt (1992).  Completeness limits were estimated by adding artificial stars 
in half magnitude intervals to the images and extracting them using DAOFIND in 
the Image Reduction and Analysis Facility
(IRAF).\footnote{IRAF is distributed by the National Optical Astronomy Observatory, which is
operated by the Association of Universities for Research in Astronomy, Inc., under cooperative agreement 
with the National Science Foundation.}
Recovery of $\ge$90\% of the artifical stars occurred for R$\le$18.1 mag and I$\le$17.0 mag.  
Photometry was not reliable for stars with R$\le$12.4 mag and I$\le$11.8 due to saturation of the CCD.  

An I vs. (R-I) diagram for over 2500 stars is presented in Figure 2.  The ordinate is the absolute I magnitude assuming a distance of 130 pc. For stars saturated in our images, R and I band photometry was adopted from the studies of Chini (1981), Gordon and Strom (1990, unpublished data), Bouvier \& Appenzeller (1992), or Walter \etal (1994).  Photometry from Chini was transformed into the Kron-Cousins system using the relations derived by Fernie (1983). The diagram is complete for the brightest stars which are candidate YSOs except for RXJ~1624.9-2459, ROXR1-4/SR-8, SR~24n, HD~148352, and ROXR2-15 for which previous photometry was not available. For comparison, PMS isochrones and the zero-age main sequence (ZAMS)
derived from the models of D'Antona \& Mazzitelli (1997) are shown (see Section 4.4).  
One of the known association members, WL~18, has strong H$\alpha$ emission (EW = 96 \AA) relative to the R band continuum 
which may explain its position below the ZAMS. 

From this diagram, we have drawn a sample for spectroscopic follow-up of stars which are on or 
above the 10$^7$ year isochrone and brighter than our completeness limits. 
Samples selected from this region of the color magnitude diagram should be representative for 
stars earlier than M6  with A$_v$ $\le$ 3 mag and ages $\le$ 10$^7$ years.           
 
Accurate positions ($<$0.5\asec) for these stars were obtained using the ASTROM program distributed by the Starlink Project and a set of secondary astrometric standards.  The secondary position references were 27 H$\alpha$ emission line stars with accurate positions determined relative to SAO stars in a 5 square degree region on the Red Palomar 
Sky Survey plate (Wilking \etal 1987). These positions were compared to counterparts from the Two Micron All Sky Survey (2MASS); matches were found within a radius of 2 \arcsec\ .  Positions were then shifted by +0.2 s in right ascension to bring them into agreement with the 2MASS coordinate system (Cutri \etal 2003).

\subsection{Hydra Observations}

Optical spectra were obtained for stars located over a 1.3 deg$^2$ area centered on L~1688 using Hydra, the multi-fiber spectrograph, on two different telescopes. Fiber configurations were designed to observe the maximum number of 
candidate YSOs; crowding at the edges of our field restricted the number of sources that could be observed. The first set of observations were made using Hydra on the Blanco 4-m telescope at Cerro Tololo Inter-American Observatory on 2003 August 10-12. 
The Bench Schmidt camera with the SiTe 2kx4k CCD gave a 40\amin\ field of view.  The 2 \arcsec\ diameter fibers
coupled with the 790 lines mm$^{-1}$ KPGLD grating yielded a wavelength coverage of 6275-7975 \AA\ centered near 7125 \AA. The spectral dispersion was 0.90 \AA\ pixel$^{-1}$ giving an effective resolution of 2.7 \AA. The resolution at the central wavelength was $\lambda$/$\Delta\lambda$=2600. The second set of observations utilized Hydra on the WIYN 3.5m telescope on 2006 June 15-17 \footnote{The WIYN Observatory is a joint facility of the University of Wisconsin-Madison, Indiana University, Yale University, and the National Optical Astronomy Observatory.}. 
The Bench Spectrograph Camera was used with the T2KA CCD which gave a 1\deg\ field of view.  The red fiber cable (2 \arcsec\ diameter) 
and the 1200 lines mm$^{-1}$ grating with a blaze angle of 28.7$\deg$ were combined with the GG-495 filter to 
cover the range of 6250-7657 \AA\ centered near 6960 \AA.  The spectral dispersion was 0.68 \AA\ pixel$^{-1}$ giving an effective resolution of 1.4 \AA. The resolution at the central wavelength was $\lambda$/$\Delta\lambda$=5000.  Three fiber configurations were observed at each telescope, set to observe 
overlapping regions of the 1.3 square degree target field. 

The spectra were reduced using IRAF.
Images were processed for bias and dark corrections using CCDPROC.  Multiple exposures of a given field were median-combined and then reduced with IRAF's DOHYDRA package. The images were flat-fielded using dome flats obtained for each fiber configuration. Sky subtraction was accomplished using the median of 7-10 sky spectra distributed 
across the field for each configuration.  
The spectra were wavelength calibrated using 5 s exposures of the PENRAY (CTIO: He, Ne, Ar, Xe) or CuAr (WIYN) lamps taken in each fiber configuration.  Scattered light corrections were not made and no flux calibration was performed.  In Table 1, we summarize the observations by presenting for each field the observation date and telescope, pointing center, number of candidate YSOs observed, number of exposures, and the total integration time.  The typical signal-to-noise ratio
for stars with R=16 mag was 30 for the CTIO spectra and 20 for the WIYN spectra as measured by line free 
regions of the continuum.

\section{Analysis of the Spectra}

As in Paper I, spectral types were derived from visual classification (visual pattern matching of our smoothed program star spectra with standard star spectra) supported by quantitative analysis of some spectral indices.  For the purposes of matching spectral features with those of standard stars, our Hydra spectra were smoothed using a Gaussian filter to the resolution of the standard stars used for direct comparison. All spectra have been normalized to 1 by dividing out a fit to the continuae, carefully excluding regions with emission lines or broad absorption due to TiO.  Normalized spectra were smoothed to a resolution of 5.7 \AA\ for comparison with the spectral standards of Allen \& Strom (1995).  For giants and later type dwarfs (M5V - M9V), optical spectra from the study of Kirkpatrick \etal (1991) were used with an effective resolution of either 8 or 18 \AA.
The relative strength of absorption due to H$\alpha$ and a blend of Ba~II, Fe~I, and Ca~I
centered at 6497 \AA\ is the most sensitive indicator of spectral type for F-K stars and the depth of the TiO bands 
for K-M stars.

A rough estimate of the surface gravity of an object is important in distinguishing PMS stars from background giants.  The primary gravity-sensitive absorption feature available for analysis in our spectra was the CaH band centered at 6975 \AA.  This band is evident in the spectra of dwarf stars with spectral types later than K5.  Following Allen (1996), we have calculated a CaH index as the ratio of the continuum at 7035$\pm$15 \AA\ to the flux in the CaH absorption band at 6975$\pm$15 \AA\ and a TiO index (primarily temperature sensitive for stars $\ge$ K5) as the ratio of the continuum at 7030$\pm$15 \AA\ to the flux in the TiO absorption band at 7140$\pm$15 \AA\ from the unsmoothed, normalized spectrum of each program object. In 
Figure 3, we plot the CaH index versus TiO index for 136 program objects.  Error bars are computed based on the one-sigma error in the mean in flux averages and propagated to the ratios.  The solid lines represent first- or second-order fits to the standard star spectra. 
 
%The remaining objects appear to have surface gravities intermediate 
%between those of giants and dwarfs but most closely resembling those of 
%dwarfs.  Therefore, spectral types for our program objects were 
%determined by comparing them to a grid of dwarf standards.

\section{Results}

Spectral types were determined for 174 of 184 stars in this study.  These data are presented in Table 2 along with any previous source names, X-ray associations, RA and decl. in J2000, the (R-I) color indices, I magnitudes, and previous spectral classifications.  The presence of lithium absorption at 6707 \AA\ and the equivalent width of H$\alpha$ are also given (emission shown as a negative value). Based on the CaH index, we have identified nine background giants and seven possible dwarfs.   Spectral classifications agree well with optical and infrared spectral types previously published by Bouvier \& Appenzeller (1992), Martin \etal (1998), Luhman \& Rieke (1999), Cieza \etal (2010),
as well as those in Paper I.  A notable exception is WLY~2-48/ISO-Oph~159.  Geers \etal (2007) report an optical spectral type of M0 while Luhman \& Rieke (1999) classify it as earlier than F3.  Our spectrum shows broad H$\alpha$ absorption partially filled in with emission as well as an absorption line from OI at 7774 \AA\ characteristic of early-type stars.  The strength of the latter plus the absence of an absorption line due to the blend at 6497 \AA\ (Ba~II, Fe ~I, Ca~I) leads us to a spectral classification of A0.  This spectral classification is in agreement with that derived by McClure \etal (2010). 

When combined with the results of Paper I, optical spectra have been
obtained for 87\% of the stars in the M(I) vs. (R-I) diagram that fell above our completeness limit
and on or above the 10$^7$ year old isochrone.  A reanalysis of the R and I band photometry has led us to revise some of the magnitudes published in Paper I.  The revised photometry is presented in Table 4 in the Appendix and used to derive the stellar parameters in this paper.

\subsection{Emission-line Spectra}

In our previous study, 39 of 131 sources (30\%) were found to have strong 
H$\alpha$ emission characteristic of CTTS.  In this sample, which was not 
biased toward the detection of H$\alpha$ emission, 15 sources were found to have 
EW(H$\alpha$) $>$ 10 \AA.  All of these are newly identified CTTS using this coarse criterion (Herbig \& Bell 1988).  
An additional 
17 objects showed weaker H$\alpha$ emission (10 \AA\ $>$ EW(H$\alpha$) $>$ 5 \AA) with all but one having an M spectral type.  The variable nature of H$\alpha$ emission is evident when comparing stars observed days apart and stars observed in this study and in Paper I.

\subsection{Identification of Pre-main-sequence Association Members}

Identification of 35 new PMS objects was accomplished using the same membership criteria 
as in Paper I with some additional criteria as described below.  
An additional 13 YSOs with optical spectral types were taken from the literature.  Combined with the 87 association members
from Paper I, there are a total of 135 objects with optical spectral types that meet one or more of these criteria.
These objects are listed in Table 3.\footnote{One object, [WMR2005] 3-17, was removed as an association member 
from Paper I since one could not rule out that it is a background giant.}   
Ninety percent of the YSOs in our sample have K or M spectral types. 

The criteria include the following:

\begin{itemize}
\item H$\alpha$ in emission with EW $>$ 10 \AA\ during at least one observation, characteristic of CTTSs.  Fifty-three objects fit this criterion.  

\item Association with X-ray emission is a signpost of youth and has been observed in 82 stars in our sample.  

\item The presence of lithium absorption is an indicator of youth for stars with spectral type K0 and cooler and clearly resolved in the spectra of 46 stars. 

\item A mid-infrared excess as observed by Infrared Space Observatory (ISO) with a spectral index from 2.2-14 $\mu$m (Bontemps \etal 2001) or the Spitzer Space Telescope from 3.6-24 $\mu$m (Wilking \etal 2008) indicative of a circumstellar disk.  Thirty-eight objects display a mid-infrared excess. 

\item A proper motion in common with the association mean as noted by Mamajek (2008).  In addition, we include Object 1-3 from this study as a common proper motion member based on data from the UCAC3 catalog. 

\item Two early-type stars, HD~147889 and Source~1, are associated with reflection nebulosity in the R and I band images. 
  
\item Finally, 100 objects (excluding giants identified in Section 3) that are too luminous to be main sequence objects at the distance to $\rho$ Oph {\it and} have an estimate for A$_v$ too high to be foreground to the cloud (A$_v$ $>$ 1.5 mag).
\end{itemize}

\noindent We have noted in Table 3 objects with near-infrared variability from the study of 
Alves de Oliveira \& Casali (2008) but do not require this for association membership.  
Only one of the objects classified as a possible dwarf based on the CaH
index, Object~4-52, displayed a criterion for association membership and is included in Table 3. 

   Mamajek (2008) has noted that proper motion data for two of the objects in Table 3, X-ray sources GY~280 
and HD~148352, are discordant and are possible foreground objects.  
Despite this fact, we include them in Table 3 but note that 
X-ray emission alone may not be sufficient to identify YSOs.  

\subsection{Distribution of Association Members}

The distribution of the 135 association members identified by this study is 
shown in Figure 1 relative to contours of $^{13}$CO column density 
which delineate the cloud boundaries (Loren 1989).  
Star symbols mark the locations of the multiple B star $\rho$ Oph and the 
three most massive members of the L~1688 embedded cluster: HD~147889, Source~1, and SR~3.  
While the association members are concentrated toward the molecular gas, 
there is marked lack of association members in the direction of the cold, 
dense cores B and E.  A comparison of this distribution with that of 
association members identified at all wavelengths (Wilking \etal 2008) 
confirms that we are missing the youngest and most highly obscured YSOs in the cloud.  

\subsection{Hertzsprung-Russell Diagram}
To derive luminosities, we began by dereddening sources using the $(R-I)$ color 
index assuming the reddening law derived by Cohen \etal (1981) in the Kron-Cousins system:

A$_v$ = 4.76 E(R-I), Sp Ty $<$ A0 (early-type) (1)

A$_v$ = 6.25 E(R-I), Sp Ty $\ge$ A0 (late-type) (2)

\noindent where R = 3.2 for early-type stars and 3.8 for late-type stars. R and I band data were taken from this study except where noted in Column 6 of Table 3. 
%R and I band data from Chini (1981) were transformed into the Kron-Cousins system using the prescription by Fernie (1983). 
For five sources, R and I band data were not available and sources were dereddened using J and H band data from the 2MASS survey (Cutri \etal 2003) and transformed into the CIT photometric system using the relationships derived by Carpenter (2001). A sixth source (WL~18) was dereddened using J and H magnitudes since strong H$\alpha$ emission distorts its R band magnitude. In these latter cases, we used the relation for the CIT photometric system

A$_v$ = 9.09 E(J-H) (Cohen \etal 1981). (3)

\noindent In a few instances, the errors in the photometry and/or spectral classifications yielded negative values for the extinction of a few tenths and an extinction of 0.0 was assumed.

Effective temperatures were derived from the spectral classifications with typical uncertainties of 
$\pm$150 K for K-M stars.  We note that the assumption of dwarf, rather than subgiant, 
surface gravities will {\it overestimate} T$_{eff}$ for 
stars with spectral types of G5-K5 by $<$250 K and {\it underestimate} T$_{eff}$ for stars with 
spectral types of M2-M8 by $<$200 K (e.g., Drilling \& Landolt 2000). 
Intrinsic colors and bolometric corrections for dwarf stars were taken from the works 
of Schmidt-Kaler (1982) for B8-K5 stars and from Bessell (1991) for K5-M7 stars. 
For the three B stars, we adopted the intrinsic colors, bolometric magnitudes, and temperatures 
from Drilling \& Landolt (2000), converting the colors into the Kron-Cousins system. 
For the M8 brown dwarf candidates, we assumed values of T$_{eff}$ = 2400, $(R-I)_0$ = 2.5, 
and BC(I) = -1.7 (Dahn \etal 2002; Hawley et al. 2002).

The absolute I magnitude, M(I), was computed given the extinction and assuming a distance of 130 pc. We then derived the bolometric magnitude and luminosity given:

M$_{bol}$ = M(I) + BC(I) (4), 

\noindent and 

log(L$_{bol}$/L$_{\sun}$) = 1.89 - 0.4 M$_{bol}$ ,  M$_{bol}$($\sun$) = 4.74. (5) 

\noindent In the situations where J and H were used to deredden, M(J) is recorded in Column 7 of Table 3
and was used along with BC(J) to derive M$_{bol}$. The median error in log(L) is computed to be 0.14 dex 
by adding quadratically errors in the R and I photometry, an uncertainty in the 
distance modulus of 0.17 mag corresponding to a depth of 10 pc, and an uncertainty 0.03 mag in the intrinsic color and 
0.1 mag in the bolometric correction 
due to spectral type errors. In most cases the dominant error was the uncertainty in the distance modulus. 

H-R diagrams for 135 association members were made using tracks and isochrones from D'Antona \& Mazzitelli (1997, DM)
F. D'Antona \& I. Mazzitelli (1998, private communication), Palla \& Stahler (1999, PS99), and Siess \etal (2000).
Despite the differences in treatments of the equation of states as a function of mass and of convection, the models give very similar results for our sample.  The diagrams for the former two are shown in Figure 4 as they represent the broadest range in mass. The masses and ages interpolated from the DM models are given in Table 3.  Since most of the objects lie on convective tracks, uncertainties in the mass {\it relative to the DM models} were estimated from the errors in the spectral classifications and uncertainties in the age from errors in the luminosities.  Uncertainties in the mass for objects in the range of 0.08-1.3 M$_{\sun}$ were typically 16\%-30\%, with the higher value corresponding to the lower mass objects. Uncertainties in the log(age) were 0.17-0.25 dex relative to DM models  with the greater uncertainty for the higher mass objects. We note that uncertainties in the absolute masses and ages
will be larger with 
theoretical mass tracks underpredicting absolute
stellar mass by 30\% - 50\% (Hillenbrand 2009).
No age or mass estimate was possible for RXJ~1624.9-2459
as it fell below the 10$^8$ year isochrone.

\subsection{Age Distribution}
The values for log(age) derived from the DM models are consistent with a normal distribution with an average log(age)
of 6.49$\pm$0.05 (3.1 Myr).  Ages derived from the PS99 models agree surprisingly well while the Siess \etal models yield
systematically older ages for log(age)$\le$7.0; the difference can be as much as 0.4 dex for a DM age of 1 Myr.
Given the large areal coverage of our survey (1.3 deg$^2$ or 6.8 pc$^2$), which must
include members of the Upper Sco subgroup, an age spread in our sample would not be unexpected.
However, simulations do not suggest an {\it intrinsic} age spread.  Assuming Gaussian-distributed errors in log T and
log L and using the DM models for 3 Myr, a Monte Carlo simulation derived values of log T and log L for over 12,000
samples in the mass range of 0.12 - 1.0 M$_{\sun}$ weighted by the Chabrier (2003) system mass function.  
While the simulated age distribution appears somewhat
narrower than what we observe, a Kolmogorov-Smirnov (K-S) test cannot reject the null hypothesis that
the two samples are drawn from the same parent
population at the 3\% level.  Lack of strong evidence for a large age spread is consistent with what is found in
other young clusters or associations (e.g., Hillenbrand 2009; Slesnick \etal 2008)
and supports the idea of rapid star formation (Hartmann 2001).
   
The average age for this sample is somewhat older than the average of 
0.3 Myr derived from more obscured sources 
in the core using the DM models and dereddened using JHK photometry 
(e.g., Greene \& Meyer 1995; Luhman \& Rieke 1999; WGM99; Natta \etal 2002).  
However, we note that there are systematic differences
in our derived luminosities, and hence ages, when (J-H) photometry is used to deredden sources instead of (R-I).
Using (J-H) colors from 2MASS yields systematically higher values for A$_v$ and hence L$_{bol}$.  As a consequence, the average
log(age) for our sample dereddened with (J-H) colors is 6.14$\pm$0.05 (1.4 Myr) compared 
to 6.49$\pm$0.05 (3.1 Myr) when (R-I) is used.
Indeed, a K-S test applied to both versions of our age distributions suggests that the difference is significant. 
The reason for this discrepancy is not understood but could involve the adopted reddening law (Cohen \etal 1981),
surface gravity effects, or possible excess emission in the J and H bands from optically-thick disks.  
J and H band excesses will overestimate the luminosity which leads to an underestimate of ages 
(Cieza \etal 2005).
Regardless, the older average age for our sample relative to the more
obscured sources is significant when both samples are dereddened using (J-H) (1.3 Myr vs. 0.3 Myr)
even considering that the previous studies used the pre-Hipparcos distance of 160 pc.
%especially considering that previous studies have underestimated the luminosity and age
%by adopting a distance of 160 pc. 

\subsection{An Extinction-limited Sample}
To explore the frequency of circumstellar disks and the IMF,
we considered 123 objects that
formed an extinction-limited sample with A$_v$$\le$ 8 mag.  This sample is representative for objects with 
M$\ge$0.2 M$_{\sun}$ for an
age of 3 Myr using the DM models. 

\subsubsection{Disk Frequency}
Using published data from the ISO (Bontemps \etal 2001) and the Spitzer Space Telescope 
(Evans \etal 2009), we can look for a mid-infrared excess relative to the photosphere 
using the spectral index 
from 2.2-14 $\mu$m or 3.6-24 $\mu$m, respectively.
The spectral index is defined as

\begin{displaymath}
\alpha = dlog\lambda F_{\lambda}/dlog\lambda
\end{displaymath}

\noindent A mid-infrared excess is defined as $\alpha$ $\ge$ -1.60 which is characteristic of an optically-thick disk 
(e.g., Greene \etal 1994).  This results in 33 of the 123 sources, or 27\% $\pm$ 5\%, showing evidence for a 
circumstellar disk lacking a large inner hole (see Table 3) 
with the uncertainty estimated assuming Poisson statistics (Gehrels 1986).  
We adopt this disk frequency estimated over 
near- to mid-infrared wavelengths as the most reliable disk indicator.
The sample size is not sufficient to investigate possible variations in the disk frequency with spectral type.

As a check, we can use the slope of the 3.6-8.0 $\mu$m flux densities from the 
Spitzer Space Telescope to assess the fraction of sources with mid-infrared excesses.  
A linear least squares fit was made to the flux densities for each source compiled from the 
Spitzer c2d catalog available in NASA/IPAC Infrared Science Archive.  
Following Lada \etal (2006), disk models suggest that $\alpha$ $\ge$ -1.80 over this wavelength range 
is indicative of an optically thick disk. The distribution of spectral indices as a function 
of spectral type is shown in Figure 5.  For the 122 sources for which data were available, 
40 or 33\% $\pm$ 5\% showed evidence for an optically thick disk consistent with our previous estimate.
We can compare this disk 
frequency to that derived by Lada \etal for the IC~348 cluster who considered 299 YSOs, 
most of which reside in area completely sampled for M $>$ 0.3 M$_{\sun}$ and 
A$_v$ $<$4 mag (Luhman \etal 2003).  Their value of 30\% $\pm$ 4\% for IC~348 is consistent 
with our disk frequency which is not surprising given the similarity in the cluster's estimated 
age of 2-3 Myr (Herbst 2008).

With knowledge of the spectral types, we can also look for evidence of even smaller infrared excesses 
from the inner disk in the K band.  Using data from the 2MASS survey, 
we transformed the magnitudes into the CIT system and dereddened them using Equation (3).  
Assuming that the excess at J was zero, the difference of the dereddened (J-H) color to the 
intrinsic color was then used to estimate 

\begin{displaymath}
r_H = F_{H_{ex}}/F_H 
\end{displaymath}

\noindent where F$_{H_{ex}}$ 
is the broadband flux from circumstellar emission and 
F$_H$ is the expected stellar flux at 1.6 $\mu$m.  This was then used to estimate the excess at K:

\begin{displaymath}
r_K = F_{K_{ex}}/F_K =(1 + r_H)(10^{[(H-K)-(H-K)_0-0.065A_v]/2.5} - 1). 
\end{displaymath}

\noindent Values of $\Delta$K = log(1 + r$_K$) were computed and those that equaled or exceeded 0.20 were 
associated with an optically-thick inner disk (Skrutskie \etal 1990).   
We note that three sources with an apparent K-band excess did not display a mid-infrared excess;
all three have values close to $\Delta$K=0.20 and may be identified as excess sources 
due to photometric errors.  Twenty-three of 123 sources, or 19\% $\pm$ 4\%, showed evidence for an 
optically thick inner disk. While the assumption of no excess emission at J could underestimate 
$\Delta$K, the higher percentage of mid-infrared excess sources 
is typical of young clusters and likely reflects the greater sensitivity of mid-infrared photometry
to disk emission (e.g., Meyer \etal 1997; Haisch \etal 2000).  
The dispersal of the inner disk by accretion, stellar winds, and/or planet formation could also 
contribute to the lower disk frequency derived from the K-band.  
Candidate transition disk objects with a mid-infrared excess and 
$\Delta$K$\le$0.10 include WSB~18, ISO-Oph~1, WSB~52, 
ISO-Oph~195, and Object~2-57, SR~21, DoAr~25, SR~9, and WMR~2-37.  The latter four have been confirmed 
as transition disk objects through
modeling of their spectral energy distributions from optical through millimeter wavelengths
(Eisner \etal 2009; Cieza \etal 2010; Andrews \etal 2011).  
 
\subsubsection{Initial Mass Function}

Previous investigations of the IMF in $\rho$ Ophiuchi have produced diverse results.
Luhman \& Rieke (1999) used K band spectra for approximately 100 stars; mass estimates for 36 plus 
completeness corrections were used to construct an IMF. Their IMF is roughly flat from 0.05-1 M$_{\sun}$ 
and peaks at $\sim$0.4 M$_{\sun}$.  de Marchi \etal (2010) use these data and fit it 
to a tapered power-law,
and derive a characteristic mass of 0.17 M$_{\sun}$.  Marsh \etal (2010) also derive an IMF for $\rho$ Ophiuchi which 
continues to rise into the brown dwarf regime, however this study was completely photometric in nature.

We divided the 123 YSOs that formed an extinction-limited sample (A$_v$$\le$ 8 mag) into mass bins 
with a width in log(mass) of 0.4 dex. No correction was made for close binaries.
A plot of the resulting mass function, shown in Figure 6, displays a peaks and turns over at 0.13 M$_{\sun}$.
\footnote {Objects in mass function with known subarcsecond companions include GSS~5, 
HD~147889, WLY~2-2, WSB~38, SR~12, SR~9, VSSG~14, SR~24N, ROX~31, SR~20, and SR~13 (see Barsony \etal 2003, 
and references therein).}  
The last three mass bins are not complete for A$_v$ = 8 mag, 
so completeness corrections were made assuming an age of 3 Myr.
To this end, the maximum visual extinction to which a source could be observed, A$_{v_{max}}$, was estimated for a 
source in the center of each mass bin using the DM models and assuming an I band limiting magnitude 
of 15.9. This value, estimated from Figure 2, is where the 3 Myr isochrone intersects
our completeness limit.
To account for variations in extinction within the region, we used the extinction maps 
from Ridge et al. (2006) with an effective resolution of 3\amin. 
The fractional area of our survey box with A$_v$=0-2, 2-4, 4-6, 6-8, 
and $>$8 mag was estimated to be 0.09, 0.32, 0.24, 0.14, and 0.21, respectively.  
We then estimated the number of missing sources for each extinction interval assuming a 
uniform stellar surface density weighted by the fractional area.  For the three lowest mass bins, 
the number of sources would increase by a factor of 1.04, 1.43, and 3.32.  
While our completeness corrections are dependent on the choice of PMS model and reddening
law, the use of other models instead of DM does not significantly change
these corrections.

For comparison with our IMF, the lognormal system mass function derived by Chabrier (2003) for field stars 
with M$\le$1.0 M$_{\sun}$ is shown in Figure 6 integrated over our mass bins and normalized to the observed
number of objects in the 0.08-0.20 M$_{\sun}$ mass bin.  For M$>$ 1.0 M$_{\sun}$, 
a power law of the form $\zeta$(log m)$\propto$m$^{-1.3}$ was used.  
Error bars plotted in Figure 6 were calculated using the methods of Gehrels (1986) and multiplied by our completeness
corrections.
In order to make an overall (large scale) comparison of the IMF with other regions and the field star IMF,
we computed the ratio of high (1 - 10 M$_{\sun}$) to low (0.1 - 1 M$_{\sun}$) mass stars.
For our sources, this ratio is R = 0.12 and 0.10 when including our completeness corrections with
an uncertainty of $\pm$0.04 calculated
using the methods of Gehrels (1986).  
Our value of R  
is in agreement with values found for $\rho$ Ophiuchi and other young embedded clusters (Meyer \etal 2000).
The ratio of high-to-low mass stars for a Chabrier system IMF is 0.16, hence we conclude
that coarsely our mass function
is consistent
with that of the field star IMF.  To make a more detailed comparison, 
a K-S test was performed over the mass ranges for which no completeness corrections were 
necessary (M$>$0.2 M$_{\sun}$).
We cannot reject the null hypothesis that the samples were drawn from the same parent population with a probability
of 40\%, suggesting that the IMF of $\rho$ Ophiuchi is not significantly different from the field star IMF over this
mass range.

When examined in more detail, there might be subtle differences between our mass function and the
field star IMF. Chabrier's (2003) lognormal IMF underestimates the number of objects in the mass bin
centered at 0.13 M$_{\sun}$.  In Chabrier (2003), a characteristic mass of 0.22 M$_{\sun}$ was derived 
for the field star IMF, however we find that our most frequently occurring mass is 0.13 M$_{\sun}$. 
Moreover we note that our lowest mass bin, which is made up of brown dwarfs, has fewer objects than predicted by the model. 
These differences may be artifacts of methods used to derive our IMF.
To quantify the possible deficit, we calculated the ratio of low mass stars (0.08 - 1 M$_{\sun}$) to brown dwarfs
($\le$ 0.08 M$_{\sun}$). Using our completeness corrected sample, we derive a ratio of 9.1 (+3.3,-2.6)
with errors calculated using methods of Gehrels (1986).
While this value is within the range found by
Andersen \etal (2008) for other star forming regions, it is higher than values derived
for $\rho$ Ophiuchi by other studies
(Geers \etal 2011; Alves de Oliveira \etal 2010). We note that both of these studies were biased
toward finding brown dwarfs and only
a small subset of their data was confirmed spectroscopically.
However as shown in Figure 6, when the errors in our completeness corrected values are taken into account,
this deficit may not be significant. 
We conclude that the turnover in our IMF is real, 
but it is uncertain if there is a real deficit of brown dwarfs compared to the field star IMF. 
%To investigate the possible deficit of brown dwarfs, we randomly sampled 20000 stars from a 
%Chabrier distribution using a Monte Carlo routine with a mass range of 0.04 M$_{\sun}$ to 10 M$_{\sun}$. 
%We compared our (non-completeness corrected) sample to the simulated sample using a K-S test which indicated that, 
%when brown dwarfs are included, we can reject the null hypothesis that these samples 
%were drawn from the same parent population. This suggestes that the deficit of brown dwarfs could be real. 
Given the difficulty of estimating completeness corrections for the brown dwarf regime, 
more sensitive spectroscopic surveys are needed to sample completely this population.
	
Finally, we compare our IMF to that derived for other star forming regions,
noting that it is 
dubious to compare directly IMFs which have been derived using different methods.
The IMF in the Taurus star-forming cloud differs from most other star-forming regions in that
it displays a higher characteristic mass ($\sim$0.8 M$_{\sun}$) as well as a corresponding deficit
of brown dwarfs (Briceno \etal 2002; Luhman 2004).
The IMFs derived for the Orion Nebula Cluster (ONC), IC~348 cluster, and Upper Scorpius 
are more similar to that in L~1688.  While all 
show a turnover at low mass (Slesnick \etal 2004; Luhman \etal 1998; 
Slesnick \etal 2008), a somewhat higher characteristic mass of approximately 
0.2 M$_{\sun}$ is reported for the ONC and IC~348.  Our peak mass of 0.13  M$_{\sun}$ is very similar 
to that reported by Slesnick, Hillenbrand, \& Carpenter (2008) for Upper Scorpius, but unlike our study, 
they report an {\it excess} of brown dwarfs relative to the field.

\section{Temporal Relationship with Upper Sco}

Given the larger data set for YSOs distributed in the low extinction regions of the L~1688 cloud, 
we revisited the relationship between star formation in this extended region (6.8 pc$^2$) to that 
in the L~1688 cloud core and in the Upper Scorpius subgroup of the Sco-Cen OB association.  
As noted in Paper I and Section 4.5, spectroscopic studies of embedded sources in the 1 pc x 2 pc centrally 
condensed core have consistently yielded ages between 0.1-1 Myr when using the D`Antona \& Mazzitelli 
tracks and isochrones, with a median age of 0.3 Myr.   
The median age for the distributed population is 
significantly older than that in the higher extinction cloud core.

In Paper I we compared our H-R diagram for 88 association members in L~1688 with that 
of the 252 members of the Upper Scorpius subgroup compiled by Preibisch \etal (2002) and noted there were 
no significant differences in age between the two samples.  However, comparisons with the Upper Scorpius 
sample are more complicated since Preibisch \etal derived temperatures and luminosities in a different manner.
For example, R and I band photometry was obtained from the UKST Schmidt plates, intrinsic colors from 
Hartigan \etal (1994), and the reddening law from Herbig (1998) plus a combination of evolutionary models 
were used (but primarily those of Palla \& Stahler 1999).  
To ease the comparison between the two samples, we compiled (J-H) photometry for sources in both samples from the 2MASS
catalog (Cutri \etal 2003), transformed it to the CIT system, and derived extinctions and luminosities 
as described in Section 4.4.  A dwarf temperature scale was used to relate spectral types in both samples to effective temperatures.  
We note that a distance of 130 pc was used for L~1688 and 145 pc for Upper Sco (de Zeeuw \etal 1999).  
The H-R diagram for the Upper Sco 
sample is shown in Figure 7 relative to the theoretical tracks and isochrones from D'Antona \& Mazzitelli.  
Ages were interpolated for sources in both samples using the DM models.   
The average log(age) for the Upper Sco sample is 6.43 (2.7 Myr) compared to 6.14 (1.4 Myr) for L~1688.  
A K-S test applied to both samples
suggests that they are not drawn from the same parent population.  Hence in this reanalysis, the low mass objects 
distributed across the L~1688 cloud appear intermediate in age between low mass stars in Upper Sco 
and YSOs embedded in the centrally condensed core. Consistent with this picture is the lower fraction of K0-M5 stars
in Upper Sco with optically-thick disks (19\%; Carpenter \etal 2006) compared to $\sim$30\% from this study.

Do the timescales involved allow the formation of the distributed population of L~1688 to be
triggered by events in Upper Sco?  
If a supernova helped power an expanding HI shell originating in Upper Sco and 
moving at $\sim$15 km s$^{-1}$ as proposed by de Geus (1992),
then in 1 Myr it would move about 15 pc and barely cover the distance in the plane of the sky between
the center of Upper Sco and L~1688.  A triggering event from Upper Sco would 
be consistent with the average age difference of
$\sim$1.3 Myr between low mass stars in Upper Sco and L~1688.  
By retracing the motions of high proper motion objects, 
Hoogerwerf \etal (2001) suggest that a supernova in a binary system
occurred in Upper Sco about 1 Myr ago that produced the runaway star $\zeta$ Oph and the pulsar PSR J1932+1059.
But this would have been too recent for a shock wave to cross the 15 pc expanse between
the two regions and initiate the formation of low mass YSOs in L~1688
with an average age of 2-3 Myr, thus requiring an earlier event.

\section{Summary}
Over 200 moderate resolution optical spectra were obtained for candidate YSOs in a 1.3 deg$^2$ area 
centered on L~1688.  When combined with the 136 spectra obtained in our initial spectroscopic study in Paper I, 
135 objects with optical spectral types
are now identified as association members based on the presence of H$\alpha$ in emission, 
X-ray emission, lithium absorption, a mid-infrared excess, a common proper motion, 
reflection nebulosity, or extinction considerations. Fifteen of these display H$\alpha$ in emission
consistent with being newly identified CTTS. 
%We find a cirumstellar disk frequency of 27\% $\pm$ 5\% for our sample, consistent with our derived age.
%Nine sources are identified as candidate transition disk objects with mid-infrared excesses and no significant K
%band excess; three of these have been confirmed by a recent multi-wavelength study.

Masses and ages were derived for association members using several theoretical models.
Using the tracks and isochrones 
from D'Antona \& Mazzitelli (1997) and F. D'Antona \& I. Mazzitelli (1998),
we derive an average age of 3.1 Myr for this distributed population. 
We find a circumstellar disk frequency of 27\% $\pm$ 5\% for our sample, consistent with our derived age and results from other studies.
Nine sources are identified as candidate transition disk objects with mid-infrared excesses and no significant K
band excess; four of these have been confirmed by recent studies.
When compared to 
simulations, our data are consistent with sampling a single age, artificially spread by uncertainties in the distance,
spectral type, and surface gravity suggesting any intrinsic age spread is small.  The age of 3.1 Myr for this surface population is 
intermediate between that of YSOs embedded in the cloud core of $\rho$ Ophiuchi and low mass stars in Upper Sco. 

We also constructed an IMF for an extinction-limited sample of 123 YSOs (A$_v$$\le$ 8 mag), 
which is a significant increase in sample size and mass range over previous studies. The resulting IMF 
is consistent with the field star IMF for YSOs with mass $>$0.2 M$_{\sun}$. 
However it may be inconsistent for masses below 0.2 M$_{\sun}$.
We find that our sample has a lower characteristic mass ($\sim$0.13 M$_{\sun}$) 
than the field star IMF as well as a possible deficit of brown dwarfs.

\acknowledgments

We are indebted to Tom Greene for offering insightful comments and suggestions.
We are very grateful to Lori Allen for assisting with the analysis of data from the Spitzer Space Telescope.
We also thank John Keller for assistance with the analysis of the Hydra data
and the NASA/Missouri Space Grant Consortium for their support.  Knut Olsen provided valuable advice 
on reduction of the CTIO Hydra data. We also thank Franceso Palla for providing a copy of the 
Palla \& Stahler models and Eric Mamajek for a program that interpolates ages and masses from the models.  
K.E. and B.W. acknowledge support from a grant from the Missouri Research Board and K.E. from a graduate fellowship
through the NASA/Missouri Space Grant Consortium.  
This research has made use of the NASA/IPAC Infrared Science Archive, which is operated by the 
Jet Propulsion Laboratory, California Institute of Technology, under contract with the 
National Aeronautics and Space Administration.

\appendix
\section{Photometry Revised from Paper I}

A reanalysis of the R and I band photometry presented in Table 2 of Paper I necessitated some revisions.  These revisions, presented in Table 4, are due in large part because of saturation problems with some of the brighter association members.  In these cases, photometry was adopted from other studies as noted.

\pagebreak

\pagebreak

\begin{center}
{\bf Figure Captions}
\end{center}
\medskip
%
% Figure 1 -
%
\figcaption{%
Distribution of association members is shown relative to contours of $^{13}$CO column density.  The contours were
computed from Loren (1989) assuming LTE and T$_{ex}$ = 25 K.   The values of the contours in units of cm$^{-2}$ are
6$\times$10$^{14}$, 3$\times$10$^{15}$, and 1.5$\times$10$^{16}$; the lowest contour delineates the outer boundary of
the dark cloud. The dashed box outlines the field included by our Hydra observations.  Star symbols mark the locations
of the star $\rho$ Oph A (labeled) and the association members Oph~S1, SR~3, and HD~147889 in the L~1688 core.}
%
% Figure 2 -
%
\figcaption{%
I vs. (R-I) color-magnitude diagram from our R and I band images.  The ordinate is the absolute I magnitude
assuming a distance of 130 pc and with no correction for reddening.  
Objects observed
spectroscopically are shown by open circles (association members), diamonds (field stars),
or "x"s (giants).  Isochrones and the ZAMS from the DM models are shown for comparison.
}
%
% Figure 3 -
%
\figcaption{%
Plot of the CaH vs. TiO indices as defined in the text for 136 program objects.  
The solid lines were derived from fits to dwarf and giant spectral standards.  For the dwarf standards from K5-M7, the fit was  y = 0.126x +0.940 with a correlation coefficient of r=0.94.
We note that for spectral types later than M7, both indices decrease in response to an overall depression of the continuum so that an M8 V star has a CaH index similar to that of an M4 V star. For the giant standards from K5-M5, the fit gave y = -0.0357x$^2$ + 0.191x + 0.795 with a correlation coefficient of r=0.82.
}
%
% Figure 4 -
%
\figcaption{%
Hertzsprung-Russell diagrams for the $\rho$ Oph association members with optically-determined spectral 
types assuming a distance of 130 pc.  The solid diamonds mark the positions of YSOs 
relative to the theoretical tracks and isochrones
of D'Antona \& Mazzitelli (1997,1998) in Figure 4a or Palla \& Stahler (1999) in Figure 4b.  
Error bars in Log T$_{eff}$ were estimated from uncertainties in the spectral type and surface gravity.
Error bars in Log L$_{bol}$ were estimated from errors in the photometry and uncertainties in the
distance modulus and bolometric correction. 
In Figure 4a, isochrones shown as solid lines are
10$^5$, 3 $\times$ 10$^5$, 10$^6$, 3 $\times$ 10$^6$, 10$^7$, and 10$^8$ years.  
Evolutionary tracks from 0.02 M$_{\sun}$ to 2.0 M$_{\sun}$ are shown by dashed lines. 
The bold dashed line marks the evolutionary track for a star at the hydrogen-burning limit. 
In Figure 4b, the birthline is shown as a solid line followed by isochrones for 10$^6$, 
3 $\times$ 10$^6$, 10$^7$, and 10$^8$ years and the ZAMS.  
Evolutionary tracks from 0.1 M$_{\sun}$ to 6.0 M$_{\sun}$ are shown by dashed lines.
}
%
% Figure 5 -
%
\figcaption{%
Spectral indices using the IRAC flux densities as a function of spectral type.  
The spectral index was computed using a linear least-squares fit to the 3.6-8.0 $\mu$m flux densities.  
Error bars were calculated from the fit given the statistical uncertainties in the flux densities. 
}
%
% Figure 6 -
%
\figcaption{%
Initial mass function for our extinction-limited sample.  Dashed lines show the members added from
the completeness corrections described in Section 4.6.2.  The dotted line is the Chabrier (2003) system mass function
which is
lognormal for M$<$1.0 M$_{\sun}$ and a power law for M$>$1.0 M$_{\sun}$.
}
%
% Figure 7 -
%
\figcaption{%
Hertzsprung-Russell diagram for 252 low-mass objects in Upper Sco from the study of 
Preibisch \& Zinnecker (1999) relative to the DM97 models.  Bolometric luminosities were derived
using (J-H) colors from the 2MASS survey (see the text).
}
\newpage
\includegraphics[angle=270]{fig1.eps}
%\plotone{fig1.eps}
\newpage
\plotone{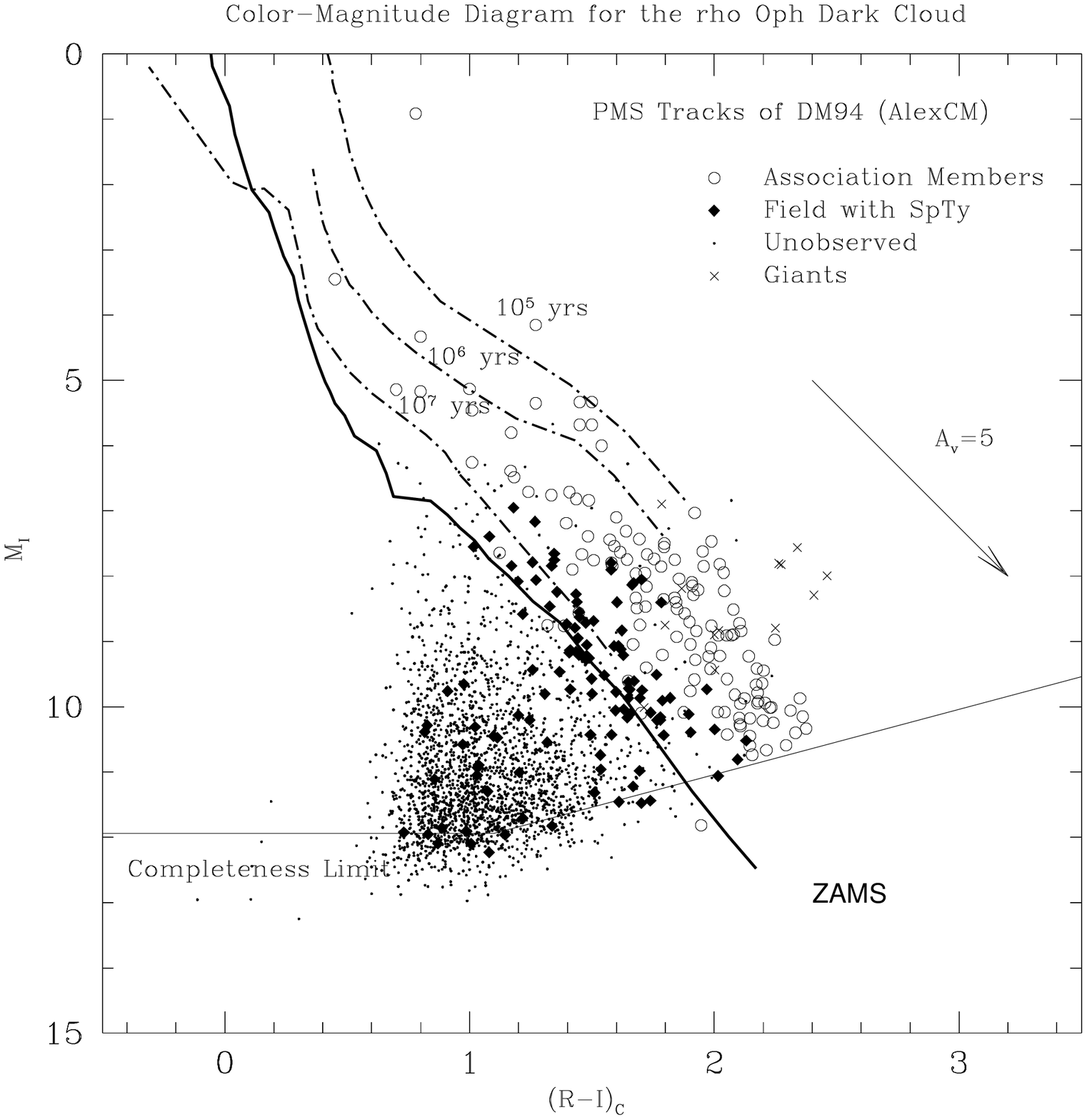}
\newpage
\plotone{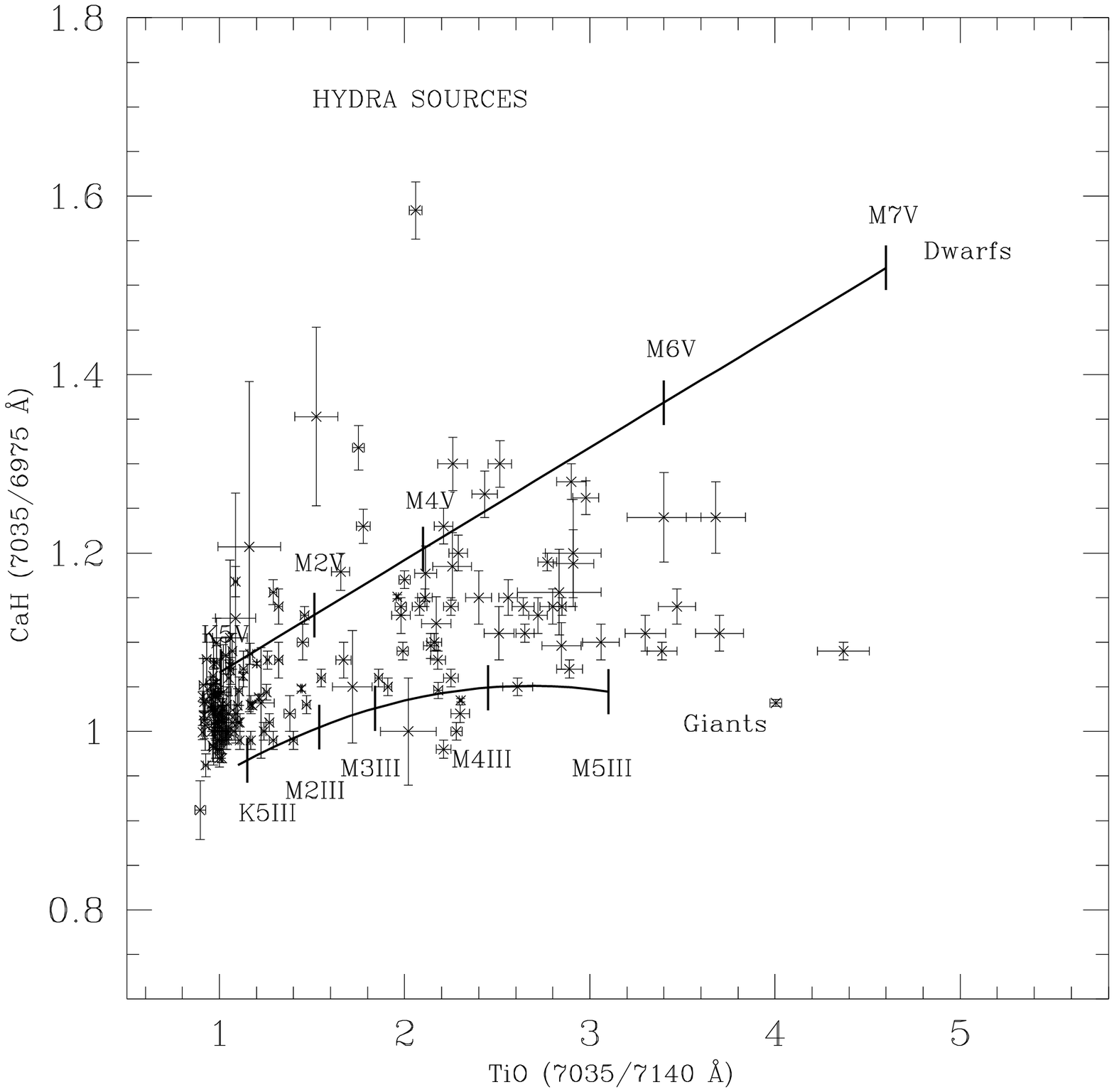}
\newpage
\plotone{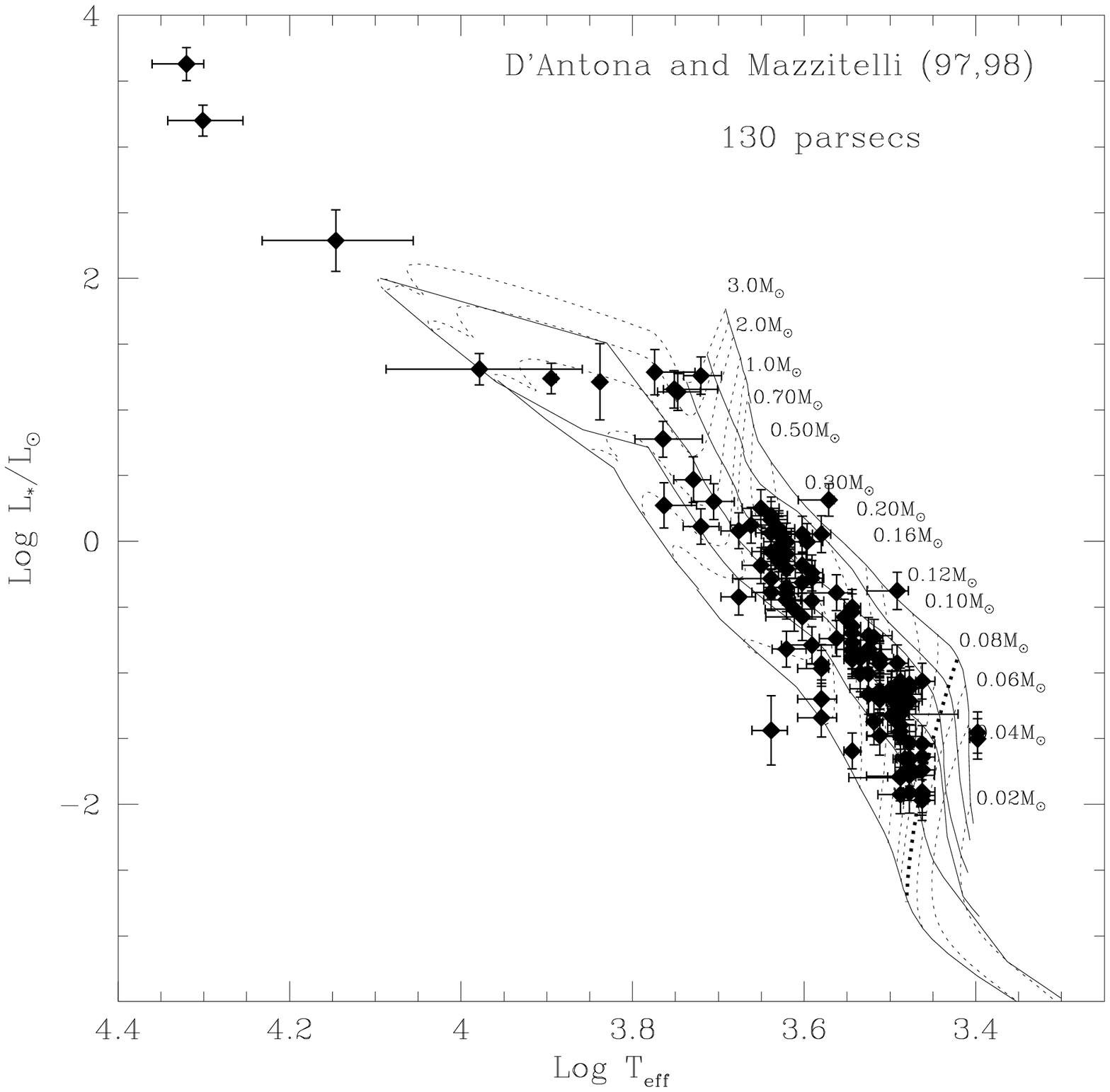}
\newpage
\plotone{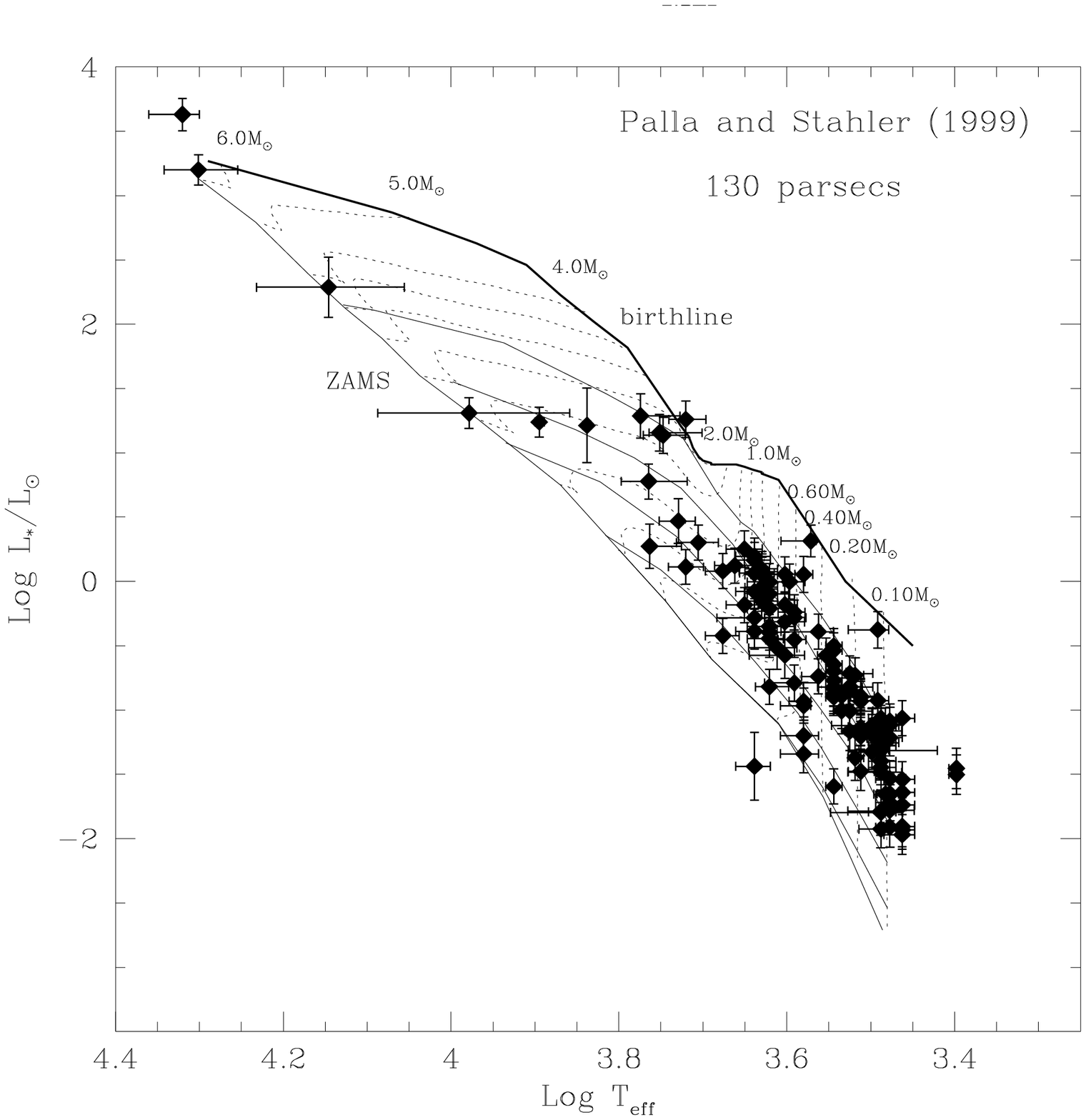}
\newpage
\plotone{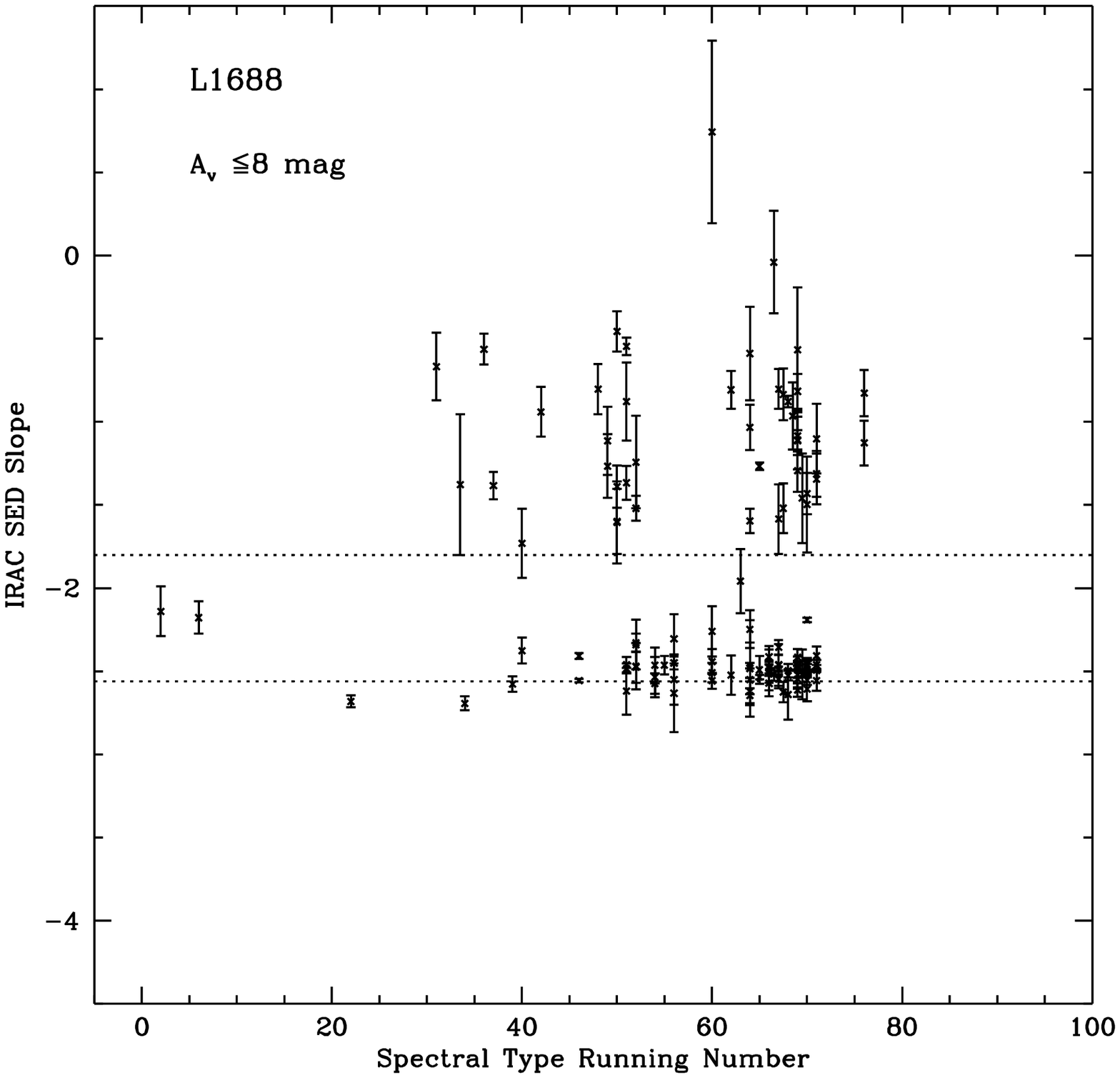}
\newpage
\plotone{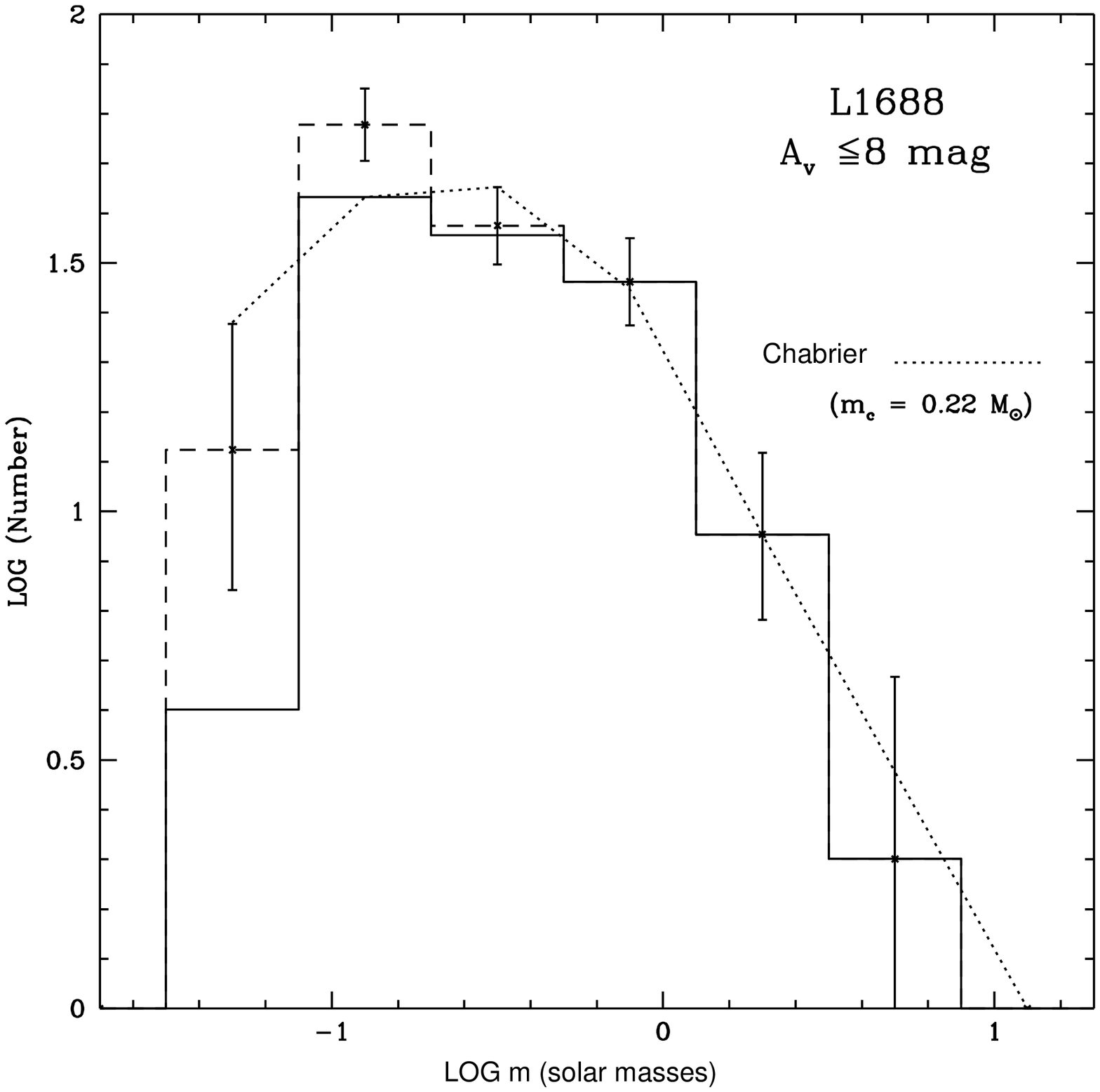}
\newpage
\plotone{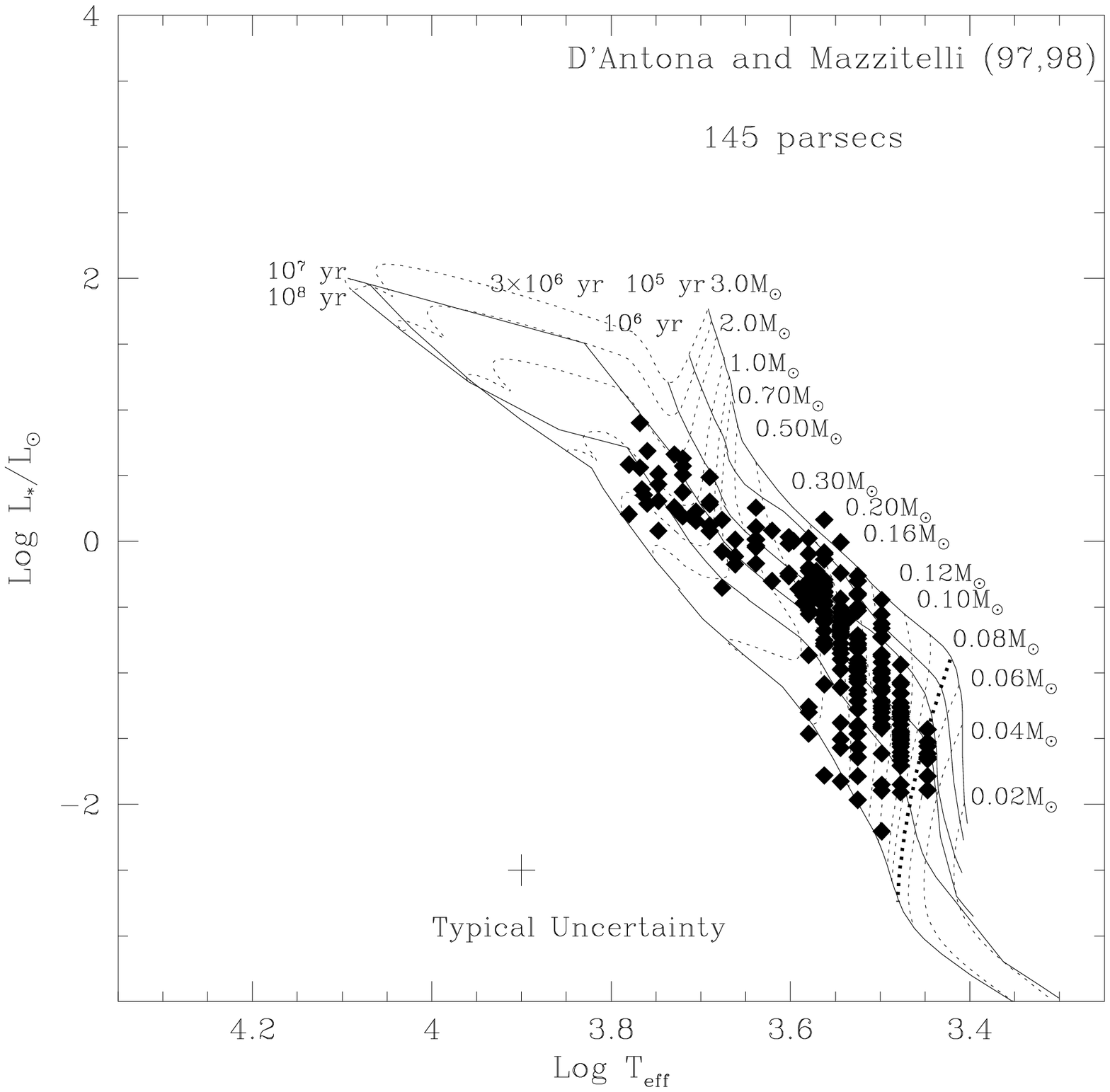}
\newpage
\clearpage
\input{tab1.tex}

\clearpage
\input{tab2.tex}

\clearpage
\input{tab3.tex}

\clearpage
\input{tab4.tex}

\end{document}

%% file: tab1.tex
%\documentclass[preprint]{aastex}
%\begin{document}
\begin{deluxetable}{ccccccc}
\tablenum{1}
\def\amin{$^\prime $}
\def\asec{$^{\prime\prime }$}
\tablewidth{0pt}
\tablecaption{Summary of Hydra Observations}
\tablehead{
\colhead{Field} & \colhead{Date} & \colhead{Telescope} & \colhead{Position} 
& \colhead{Sources} 
&\colhead{No. of Exp.} &  \colhead{Int. Time}  \\
\colhead{No.} & \colhead{(YYMMDD)} & & \colhead{(J2000.0)} & 
&  & \colhead{(minutes)} 
} 
\startdata
1  & 030810   & Blanco & 16:27:05.4  -24:45:09 & 24 & 5 & 135    \\
2  & 030811   & Blanco & 16:28:15.2  -24:14:04 & 47 & 5 & 135    \\
3  & 030812   & Blanco & 16:26:01.9  -24:14:23 & 48 & 5 & 135    \\
4  & 060615   & WIYN   & 16:25:47.2  -24:41:00 & 32 & 10& 240    \\
5  & 060616   & WIYN   & 16:28:21.6  -24:28:00 & 42 & 10& 240    \\
6  & 060617   & WIYN   & 16:27:22.9  -24:09:00 & 30 & 10& 240    \\
\enddata
\end{deluxetable}

%\end{document}

%% file: tab2.tex
%\documentclass[preprint2]{aastex}
%\begin{document}
%This is Table 2 of our HYDRA paper
%\pagestyle{empty}
%\voffset=2.4truein
\begin{deluxetable}{cccccccccccccc}
\singlespace
\rotate
\tablenum{2}
\tablecolumns{17}
\tabletypesize{\scriptsize}
\def\asec{$^{\prime\prime }$}
\def\deg{\hbox{$^\circ$}}
\def\amin{$^\prime $}
\def\etal{{\it et al. }}
\tablewidth{750pt}
\tablecaption{Optical Properties of Candidate Young Stellar Objects}
\tablehead{
\colhead{F} & \colhead{Ap.\tablenotemark{a}}
& \colhead{Names(s)\tablenotemark{b}} & \colhead{X-Ray ID\tablenotemark{c}}
& \colhead{RA(J2000)}& \colhead{decl.(J2000)}
& \colhead{Li?}
&  \colhead{EW(H$\alpha$)}
&  \colhead{I} & \colhead{(R-I)\tablenotemark{d}}
& \colhead{Prev Sp. Ty.\tablenotemark{e}}  & \colhead{Sp. Ty.} & \colhead{Adopt}
& \colhead{Notes} \\
   &   &   &  & \colhead{(hhmmss.s)} & \colhead{(\deg\ \amin\ \asec\ )} &  & 
&  & \colhead{(mag)} & \colhead{(mag)}  & & &  \\ 
}
\startdata
   3&        39&                        &                       &        16      24    39.3    & -24      07      10&  \nodata &            1.1&     15.31&      1.41&  \nodata          &  K0-K2    &  K1          &          \\
   4&        40&                        &                       &        16      24    39.7    & -24      35      08&  \nodata &           -9.2&     15.87&      2.11&  \nodata          &  M4.5     &  M4.5        &          \\
   4&        13&                        &                       &        16      24    40.5    & -24      43      09&  \nodata &  $<$0.3       &     15.96&      1.90&  \nodata          &  G5-K6    &  K0:         &          \\
   3&        11&                        &                       &        16      24    40.7    & -24      21      40&  \nodata &            0.90&     14.83&      1.49&  \nodata          &  K0-K2    &  K1          &          \\
   4&        29&                        &                       &        16      24    41.1    & -24      17      49&  \nodata &  $>$ -0.4     &     15.15&      2.05&  \nodata          &  M1-M3    &  M2          &          \\
   4&        99&                        &                       &        16      24    43.9    & -24      47      53&  \nodata &           0.70&     14.64&      1.61&  \nodata          &  K3       &  K3          &          \\
   4&        14&                        &                       &        16      24    44.8    & -25      00      18&  \nodata &          -3.5&     15.74&      1.64&  \nodata          &  M2-M4    &  M2          &  Poss. dwarf        \\
   3&        37&                        &                       &        16      24    46.9    & -24      22      21&  \nodata &            0.90&     13.61&      1.86&  \nodata          &  K6-M0    &  K7          &          \\
\enddata
\tablenotetext{a}{HYDRA field and aperture number of observation.} 
\tablenotetext{b}{Sources names from optical or infrared studies by
(GSS, Source 2) Grasdalen \etal 1973;
(VSSG) Vrba \etal 1975; (VSS) Vrba \etal 1976; 
(WSB) Wilking \etal 1987; 
(WLY) Wilking \etal 1989; 
(GY) Greene \& Young 1992; (ISO-Oph) Bontemps \etal 2001; (AOC) Alves de Oliveira \& Casali 2008;
and (UCAC3) Zacharias \etal 2009.
}
\tablenotetext{c}{X-ray source association from the EINSTEIN survey by
(ROX) Montemerle \etal (1983), the ROSAT surveys by (RXJ) Mart\'{\i}n \etal (1998) and (ROXRF) Grosso \etal 
(2000), 
the Chandra surveys by ([IKT2001]) Imanishi\etal (2001) and (ROXC) N. Grosso (2005, private communication ), 
or the XMM-Newton survey by
(ROXN) Ozawa \etal (2005).} 
\tablenotetext{d}{R and I band photometry from this study except for ROX~3 which are from 
Gordon \& Strom (1990, unpublished data).}
\tablenotetext{e}{References for spectral types: 
(BA) Bouvier \& Appenzeller 1992; (MMGC) Mart\'{\i}n \etal 1998; (LR) Luhman \& Rieke 1999;
(WMR) Wilking \etal 2005; (G) Geers \etal 2007.
}
\end{deluxetable}

%\end{document}

%% file: tab3.tex
%\documentclass[preprint2]{aastex}
%\begin{document}
%This is Table 3 of our HYDRA paper
%\pagestyle{empty}
%\voffset=2.4truein
\begin{deluxetable}{ccccccccccccc}
\singlespace
\rotate
\tablenum{3}
\tablecolumns{13}
\tabletypesize{\scriptsize}
\def\asec{$^{\prime\prime }$}
\def\deg{\hbox{$^\circ$}}
\def\amin{$^\prime $}
\def\etal{{\it et al. }}
\tablewidth{750pt}
\tablecaption{Association Members with Optical Spectra}
\tablehead{
\colhead{F} & \colhead{Ap.}
& \colhead{Names(s)\tablenotemark{a}} & \colhead{WMR\tablenotemark{b}}
& \colhead{Sp. Type\tablenotemark{c}}& \colhead{A$_v$\tablenotemark{d}}
& \colhead{M(I)}
&  \colhead{log T$_{eff}$}
&  \colhead{log(L/L$_{\sun}$)} & \colhead{M$_*$\tablenotemark{e}}
& \colhead{Log(age)}  & \colhead{Criteria\tablenotemark{f}} & \colhead{Notes\tablenotemark{g}} \\
 & & & & & (mag) & (mag) & (K) & & (M$_{\sun}$) & (yr) \\ 
}
\startdata
   4&   40&                     &       &   M4.5        &         2.1   &        9.03    &     3.488    &     -1.79  & 0.15      & 7.18&    ext  \\
   4&   29&                     &       &   M2          &         5.6   &        6.23    &     3.544    &     -0.82  & 0.32      & 6.46&    ext \\
   3&   37&                     &       &   K7          &         6.8   &        3.96    &     3.602    &     0.06  & 0.42      & 5.65&    ext \\
   4&   42&                     &       &   M4.5        &         2.3   &        7.83    &     3.488    &     -1.31  & 0.16      & 6.57&    ext \\
    &     & RXJ~1624.9-2459     &       &   K5 (MMGC)   &     0.3 (2M)  &         6.91   &     3.638    &     -1.44  & \nodata     &   \nodata    &    pm,x &(1) \\
   3&   58&                     &  2-15 &   M3.5        &         2.6   &        6.60    &     3.512    &     -0.90  & 0.19      &6.23 &    ext,li \\
   4&   59&  WSB~18             &       &   M3.5        &         2.7   &        6.58    &     3.512    &     -0.89  & 0.19      &6.23 &    ext,ha,IRX & (4) \\
   4&   82&  WSB~19             &       &   M4.5        &         0.2   &        7.46    &     3.488    &     -1.16  & 0.16      &6.38 &    ha,IRX \\
\enddata
\tablenotetext{a}{Sources names from optical or infrared studies noted in Table 2 plus:
(SR) Struve \& Rudkjobing 1949; (DoAr) Dolidze \& Arakelyan 1959; (EL) Elias 1978;
(Chini) Chini 1981; (WL) Wilking \& Lada 1983; and (ROXR1) Casanova \etal 1995.
}
\tablenotetext{b}{Source name from Wilking \etal (2005).}
\tablenotetext{c}{Optical spectral types from this study or Wilking \etal (2005) except where noted:
(EL) Elias 1978; (CK) Cohen \& Kuhi 1979; 
(HS) Houk \& Smith-Moore 1988; (GS) Gordon \& Strom 1990, private communication; (BA) Bouvier \& Appenzeller 1992; 
(TQ) Torres \etal 2006; and (MMGC) Mart\'{\i}n \etal 1998.}
\tablenotetext{d}{R and I band magnitudes used to compute A$_v$ and L$_{bol}$ are from this study except where noted: 
(C) Chini 1981;
(GS) Gordon \& Strom 1990, private communication; (BA) Bouvier \& Appenzeller 1992; and (W) Walter \etal 1994.  (2M) denotes J and H 
band data from 2MASS were used to compute A$_v$ and L$_{bol}$ and M(J) replaces M(I) 
in Column 7.}
\tablenotetext{e}{Masses and ages estimated from the tracks and isochrones of D'Antona \& Mazzitelli (1997)
and F. D'Antona \& I. Mazzitelli (1998, private communication)
except where noted in the last column.}
\tablenotetext{f}{Association membership established through location above the main sequence and A$_v$ $>$ 1.5 mag (ext),
the presence of H$\alpha$ emission EW $>$ 10 \AA\ (ha), an infrared excess (IRX), lithium absorption (li),
proper motion (pm), reflection nebulosity (ref. neb.), near-infrared variability (var), or
X-ray emission (x).  See the text for details.}
\tablenotetext{g} {Notes: (1) sits below main sequence, possible foreground star; (2) spectroscopic binary 
Casassus \etal 2008; (3) mass and age derived from the tracks and isochrones of Palla \& Stahler 1999; 
(4) subarcsecond companion observed in infrared and treated as a single object (Barsony \etal
2003, Prato 2007);
(5) discordant proper motion (Mamajek 2008).}  
\end{deluxetable}
%\end{document}

%% file: tab4.tex
%\documentclass[preprint]{aastex}
%\begin{document}
\begin{deluxetable}{lccc}
\tablenum{4}
\def\amin{$^\prime $}
\def\asec{$^{\prime\prime }$}
\tablewidth{0pt}
\tablecaption{Revised Photometry from Paper I}
\tablehead{
\colhead{Name} & \colhead{I} & \colhead{(R-I)} & \colhead{Reference} \\
} 
\startdata
$[WMR~2005] 1-29$   & 10.7    & 1.0    & Bouvier \& Appenzeller (1992)    \\
$[WMR~2005] 1-30$   & 11.57   & 1.54   & Gordon \& Strom (1990, private communication)    \\
$[WMR~2005] 1-28$   & 11.37   & 1.17   & Gordon \& Strom (1990, private communication)   \\
$[WMR~2005] 4-66$   & 17.85   & \nodata& This study    \\
$EM* SR~12$         & 10.92   & 1.27   & Gordon \& Strom (1990, private communication) \\
$[WMR~2005] 1-26$   & 9.9     & 0.8    & Bouvier \& Appenzeller (1992)    \\
$[WMR~2005] 3-19$   & \nodata & \nodata& Edge of image: this study    \\
$[WMR~2005] 1-35$   & 11.25   & 1.45   & Gordon \& Strom (1990, private communication) \\
$[WMR~2005] 1-15$   & 10.9    & 1.45   & Bouvier \& Appenzeller (1992) \\
$[WMR~2005] 4-56$   & 17.04   & 1.70   & This study\\
$[WMR~2005] 1-3$    & \nodata &\nodata & Saturated: this study \\
\enddata
\end{deluxetable}

%\end{document}